\documentclass[]{jfm}

\usepackage{graphicx}
\usepackage{newtxtext}
\usepackage{newtxmath}
\usepackage{natbib}
\usepackage{hyperref}
\usepackage{soul}
\usepackage[dvipsnames]{xcolor}
\hypersetup{colorlinks = true, allcolors=BrickRed}

\newcommand{\RomanNumeralCaps}[1]
\linenumbers

\newcommand{\dt}[1]{\mathrm{d}#1}

\title{Spatial evolution of the turbulent/turbulent interface geometry in a cylinder wake}

\author{Jiangang Chen\aff{1}
 \and Oliver R. H. Buxton\aff{1} \corresp{\email{o.buxton@imperial.ac.uk}} }

\affiliation{\aff{1}Department of Aeronautics, Imperial College London, South
Kensington Campus, London SW7 2AZ, UK}

\begin{document}
\maketitle

\begin{abstract}
This study aims to examine the spatial evolution of the geometrical features of
the turbulent/turbulent interface (TTI) in a cylinder wake. The wake is exposed
to various turbulent backgrounds in which the turbulence intensity and the
integral length scale are independently varied and comparisons to a
turbulent/non-turbulent interface (TNTI) are drawn. The turbulent wake was
marked with a high-Schmidt-number ($Sc$) scalar and a planar laser induced
fluorescence (PLIF) experiment was carried out to capture the interface between
the wake and the ambient flow from $x/d$ = 5 to 40 where $x$ is the streamwise
coordinate from the centre of the cylinder and $d$ is the cylinder's diameter.  
It is found that the TTI generally spreads faster toward the ambient flow than
the TNTI. A transition region of the interfaces' spreading is found at $x/d
\approx 15$, after which the interfaces propagate at a slower rate than
previously (upstream) and the mean interface positions of both TNTI and TTI
scale with the local wake half-width. The location of both the TNTI and TTI have
non-Gaussian probability density functions (PDFs) in the near wake because of
the influence of the large-scale coherent motions present within the flow.
Further downstream, after the large-scale coherent motions have dissipated, the
TNTI position PDF does become Gaussian. For the first time we explore the
spatial variation of the ``roughness'' of the TTI, quantified via the fractal
dimension, from near field to far field. The length scale in the background flow
has a profound effect on the TTI fractal dimension in the near wake, whilst the
turbulence intensity only becomes important for the fractal dimension farther
downstream. 
\end{abstract}



\section{Introduction}
\label{sec:intro}
All turbulent flows embedded within a non-turbulent background are observed to
spread out into their environment. The spreading of turbulence into previously
irrotational fluid depends, in the first instance, on viscous diffusion of
vorticity across a well-defined thin layer which bounds the turbulent region and
separates it from the outer, non-turbulent regions \citep{Townsend1976}. This
convoluted thin layer, usually referred to as a turbulent/non-turbulent
interface (TNTI), was first examined in detail by \citet{corrsin1955}, and
extensive studies on the dynamical and geometrical features of TNTIs in various
turbulent shear flows have been conducted ever since \citep[see the review
of][]{da2014interfacial}. However, numerous situations for
turbulent industrial and environmental flows have a turbulent background which
has a profound influence in the dynamics of the turbulence in the main stream
\citep[][]{rind2012direct, rind2012effects, pal2015effect}; a typical example is
the wake of a wind turbine developing in the atmospheric turbulent boundary
layer or the turbulent wake of other upstream wind turbines
\citep[e.g.][]{porte2020wind}. 

In contrast to the extensive studies of TNTIs, our knowledge of the interface
between flow regions with different levels of turbulence intensity, hereinafter
referred to as a turbulent/turbulent interface (TTI), remains limited,
notwithstanding its prevalence in the physical world. In the recent study of
\citet{kankanwadi2020turbulent} the entrainment across a TTI between a turbulent
cylinder-wake and a grid-generated turbulent background was experimentally
examined. The cylinder's wake was marked with a fluorescent dye
of high Schmidt number ($Sc$) such that molecular diffusion occurred at a
vanishingly small length scale  \citep[e.g.][]{watanabe2015turbulent}. By
examining the velocity field in the vicinity of the scalar-marked interface it
was revealed that a clear interface existed between the wake and the turbulent
ambient fluid, independently of the artificially-introduced scalar. In
particular, a jump in vorticity magnitude over a short distance was reported,
resembling the vorticity jump across a TNTI \citep{da2014interfacial}. Both the
intensity and the integral length scale of the background turbulence were
independently varied and it was shown that in this far-wake region the
turbulence intensity was the important parameter in determining the geometry of
the TTI, characterised by its tortuosity and fractal dimension.

In their subsequent study of the flow physics governing the behaviour of the
TTI, namely consideration of the various terms of the enstrophy transport
equation, \citet{kankanwadi2022physical}  found the magnitude of the viscous
diffusion term is insignificant when compared to that of the inertial vorticity
stretching term acting at the outermost boundary of the TTI. These results imply
that viscous diffusion is of little importance to the entrainment process across
a TTI which contrasts to the scenario of the TNTI in which viscous diffusion is
the dominant process by which the irrotational fluid acquires vorticity in the
so-called viscous superlayer \citep[e.g.][]{corrsin1955, da2014interfacial}.
\citet{kankanwadi2022physical} also demonstrated that the vorticity in the
vicinity of the TTI is ``organised'' in such a way on the wake side of the TTI
that it exploits the enhanced strain rates in the interface-normal direction,
previously reported for TNTIs
\citep[e.g.][]{buxton2019importance,cimarelli2015spectral}, thereby enhancing
vorticity stretching/enstrophy production and yielding the enstrophy jump across
the TTI.

In spite of these dynamical differences between the TTI and TNTI, their
geometries both display a common hierarchy of self-similar structures which can
be described through fractal analysis. The fractal nature of the interface
geometry, which renders a much larger surface area of the interface than
otherwise, is essential to correctly modelling the turbulent entrainment rate
\citep[e.g.][]{sreenivasan1989mixing,zhou2017related}. \citet{kohan2022scalar}
investigated the effect of the background turbulence intensity on the geometry
of the TTI of an axisymmetric jet and compared it with a TNTI. They found that
the turbulence in the ambient flow can further stretch and corrugate the
interface and thus result in a larger fractal dimension of the TTI than the TNTI
in results that corroborated those of \citet{kankanwadi2020turbulent} for a turbulent wake. It is
noted that their investigation was carried out in the far field of the jet (25
diameters downstream of the orifice) where the coherent motions of the jet have
dwindled \citep[][]{tennekes1972first, gordeyev2000coherent}. In such a
situation, the turbulence intensity in the background flow is the dominant
parameter in modifying the behaviour of the TTI, whilst the size of the
energetic eddies in the background flow, characterized by the integral length
scale, is of less relevance \citep[e.g.][]{kankanwadi2020turbulent}. 

However, when it comes to the flow region where the coherent motions prevail,
the scenario is quite different. It has been reported that the entrainment
becomes dominated by large-scale engulfment of background fluid under the
influence of the coherent motions
\citep[e.g.][]{yule1978large,bisset2002turbulent,
cimarelli2021numerical,long2022universal}. For TTIs, \citet{kankanwadi2022near}
observed that both the turbulence intensity and the integral length scale in the
ambient flow correlate to enhanced entrainment in the presence of the
large-scale coherent vortices in the near wake of a cylinder; a contrasting
result to the far-field study in which background turbulence was observed to
suppress entrainment rate \citep{kankanwadi2020turbulent}. By conducting a
control experiment in which the large-scale coherent vortices in the wake (the
von K\'{a}rm\'{a}n vortex street) were suppressed via the addition of a splitter
plate they deduced that the presence of freestream turbulence effectively
enhances entrainment via engulfment but suppresses the small-scale ``nibbling''.
\citet{kankanwadi2022near} also reported that the presence of freestream
turbulence increases the locus of the wake's large-scale coherent vortices (i.e.
wake ``meandering'' with a larger amplitude), with the integral length scale of
the background turbulence playing the most important role in determining this.
Combined, these results highlight the important role that the presence of the
large-scale coherent motions of the wake, and their interaction with any
background turbulence present, play in modulating the properties of the TTI.

The major motivation of the current study is that the previous
papers, which focus on the TTI properties in the far wake ($x/d = 40$ in
\citet{kankanwadi2020turbulent}) and the near wake ($x/d \le 10$ in \citet{kankanwadi2022near}), found essentially the opposite effect of the background
turbulence intensity on the entrainment rate into the wake. As stated in
\citet{kankanwadi2022near}, this implies the existence of a `transition'
region, in which the entrainment rate (and hence behaviour of the TTI) adjusts
from one regime to the other. Such observations raise several questions with
regard to the spatial evolution of the properties of the TTI, as the coherent
vortices degrade downstream:  
where does this `transition' happen? How do the properties of
        the TTI/TNTI, such as their location-PDF, scaling, and fractal dimension
        evolve from the near to the far field in the context of the coherent
        motions of the wake diminishing? Particularly, which parameter in the
        background turbulence dominates the local fractal dimension of the TTI,
        the intensity level of the background turbulence or the size of the
        energetic eddies? Will the dominant parameter change before and after
        the transition region? All these intriguing questions emerging from the
        findings of \citet{kankanwadi2020turbulent} and
        \citet{kankanwadi2022near} are imperative for understanding the dynamics
        of TTIs as the wake develops downstream. We aim to answer all
        these questions in the present study, so as to contribute to an in-depth
        and complete comprehension of the behaviours of the TTI/TNTI in planar
        wakes.

In order to addresses these questions, we examined the wake of a circular
cylinder in various turbulent freestreams, in which the turbulence intensity and
integral length scales of the background turbulence were independently varied. A
planar laser induced fluorescence (PLIF) experiment was conducted to capture the
position of the interface between the wake and the freestream from 5 to 40
cylinder diameters downstream from the cylinder's centre. In such a region of
the flow the coherent vortices in the wake emanating from the shear layers shed
from the cylinder experienced a significant decay
\citep{matsumura1993momentum,chen2016three}, which allows us to investigate the
streamwise evolution of both the TTI and TNTI position/geometry concerning the
questions raised above. The paper is organized as follows. Section 2 describes
the experimental details, and the visualisation of the flow and the methodology
used to determine the interface position is presented in section 3. Major
results are discussed in section 4 and we summarise and conclude the work in
section 5.

\section{Experimental setup}
\begin{figure}
    \centering
    \includegraphics[width = \textwidth]{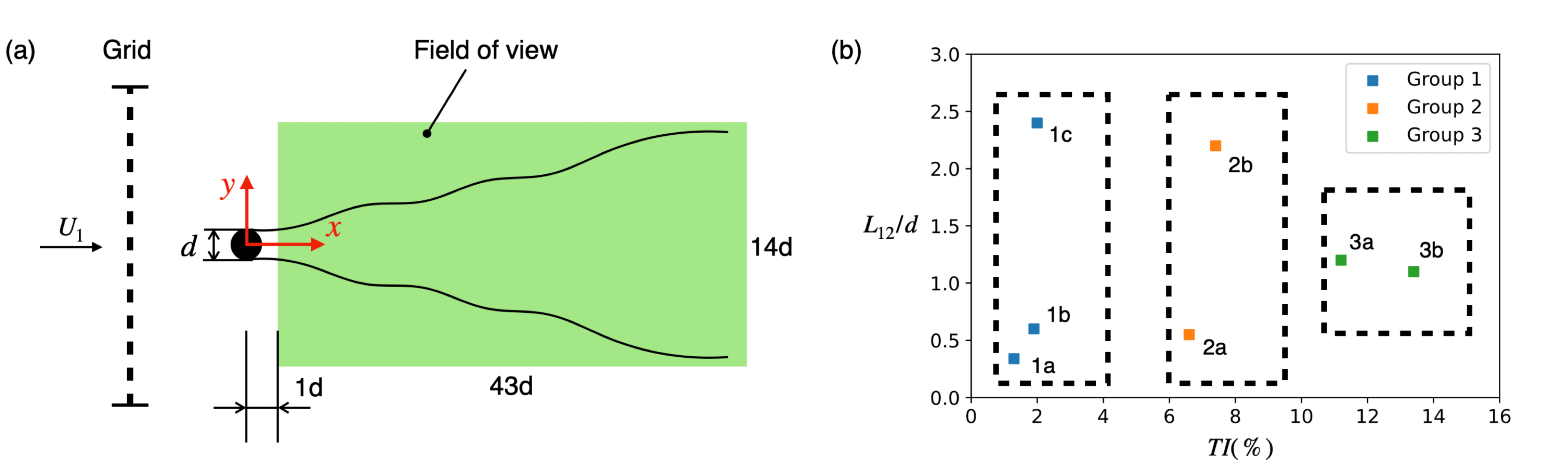}
    \caption{(a) Conceptual sketch of the experimental setup and (b) parameter
    space ($TI, L_{12}$) of the background flow in the middle of the field of
    view at $x/d = 20$.}
    \label{fig1}
\end{figure}

The experiments were conducted in the water flume of the hydrodynamics
laboratory of the Aeronautics Department at Imperial College London. A cylinder
with a diameter of $d = 0.01$ m is vertically mounted in the middle of the flume
test section which has a dimension of 9m in length and 0.6 m in cross section
which was filled to a depth of 0.6 m.  
The incoming velocity of the flow is $U_1$ = 0.38 m/s. The Reynolds number based
on $U_1$ and $d$ is about 3800. Upstream of the cylinder, four different grids,
including two regular and two fractal grids (see \citet{kankanwadi2020turbulent}
for details of the grids), are used to generate the background turbulence with
various turbulence intensities and length scales. 

A planar laser induced fluorescence (PLIF) experiment was carried out to capture
the boundary of the cylinder's wake in the various background flows. A
fluorescent dye, Rhodamine 6G, which can be treated as a passive scalar in
the flow was utilized to demarcate the wake region of the cylinder from the
background flow. For a TTI, it is impossible, to reliably
identify the interface between different turbulent regions based on vorticity
since high-magnitude vorticities appear on both sides of the interface, which is
in contrast to the situation for a TNTI. Nevertheless it has been proven that
there is a discontinuity in turbulent properties across this interface
\citep{kankanwadi2020turbulent} independently of the
(artificially-introduced) scalar. The very
high Schmidt number (Sc) of the dye, approximately 2500 in water
\citep{vanderwel2014accuracy}, ensures that the molecular diffusion of the dye
occurs over a  negligibly short length scale with respect to the turbulent
motions, so that the dye acts as a near-perfect marker of the wake region, with a
clear boundary. The dye was released into the wake from a hole in the rear
surface of the cylinder with the aid of a micro-dosing pump (B\"{u}rkert 7615)
working at a dosing frequency of 10 Hz. A long elastic tube of 2 m was used in
the routing of the dye from the pump to the hole on the cylinder so as to smooth
out pulsations in the dye release and the scalar can be fully
smeared into the wake by the turbulent motions ahead of the downstream
measurement positions, as verified in \citet{kankanwadi2020turbulent}.

\begin{figure}
    \centering
    \includegraphics[width = \textwidth]{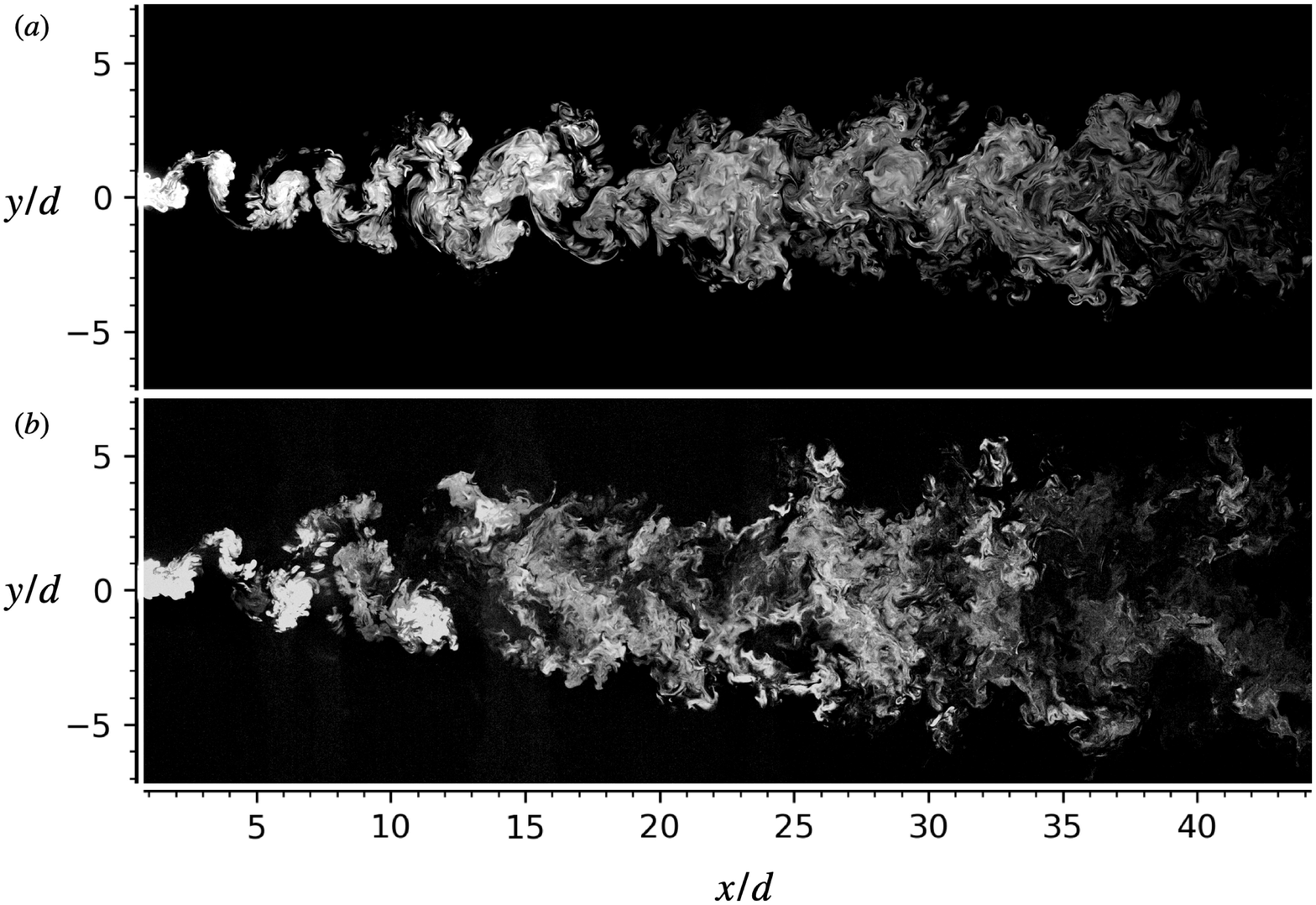}
    \caption{Visualisation of the wake (a) without 
    (case 1a) and (b) with (case 2a) turbulence present in the background flow.}
    \label{fig2}
\end{figure}

A high-speed Nd:YLF laser (Litron LDY304) with a wavelength of 527 nm was used
to induce the fluorescence of the dye which emits light of wavelength around 560
nm. The fluorescence was captured by two cameras (Phantom V641 with a sensor
resolution of $2560 \times 1600$ pixels) which were arranged consecutively in the
streamwise direction to form a field view of $14d \times 43d$ with an overlap
region of about $2.5d$. The spatial resolution of the measurement is about 0.1
mm per pixel. The typical Kolmogorov length scale at a similar Reynolds
number is about 0.16 mm \citep{kankanwadi2020turbulent}, so the the spatial
resolution is approximately 0.6$\eta$/pixel. The Batchelor scale $\eta_B$
($\equiv \eta/Sc^{1/2})$ is about $\eta/50 $. Accordingly
the smallest details of the scalar interface are negligible in size in
comparison to those of the vorticity interface, hence there is (deliberately) no
need for us to resolve the Batchelor scale in our study. The upstream edge of
the field of view is $1d$ apart from the centre of the cylinder (figure
\ref{fig1}a). A low-pass filter is placed in front of the camera lens in order
to ignore any laser light noise in the PLIF image. Instantaneous images of the
wake in a freestream without (a) and with (b) turbulence is displayed in figure
\ref{fig2}. The acquisition frequency of the experiment is 100 Hz and 2000
images were captured for each measurement case. 

Following \citet{kankanwadi2020turbulent}, we employed turbulence intensity ($TI
\equiv \sqrt{(u^2 + v^2)/2}/U_1$ where $u$ and $v$ are velocity fluctuations in the
$x$ and $y$ directions respectively) and integral length scale ($L_{12} \equiv
\int_0^{r_0} R_{12}(r)dr$ where $R_{12}(r)$ is the correlation coefficient
between $u(x,y)$ and $u(x, y+r)$) to characterize the various turbulent
background flows, and $r_0$ is the location at which $R_{12}$ first crosses zero. The distribution of the turbulence intensity and the length
scale of the flow behind the grids has been documented in detail in
\citet{kankanwadi2022turbulent} in the same facility and operating conditions.
The cylinder is placed at various downstream distances from the various grids
such that the parameter space ($TI$, $L_{12}$) was explored as widely as
possible in order to truly investigate the behaviour of the interface between
the wake and the background flow with various ``flavours'' of turbulence. We
conducted experiments for seven cases of ($TI$, $L_{12}$) and the distribution
of ($TI$, $L_{12}$) at $x/d = 20$, i.e. the middle of the field of view, is
shown in figure \ref{fig1}b. We divided the seven cases into three groups
(figure \ref{fig1}b) according to the magnitude of the turbulence intensity.
Case 1a is the closest experimental approximation to a TNTI-case with no
turbulence-generating grid mounted upstream of the cylinder. The remaining cases
are TTI cases with turbulent backgrounds generated by the four different grids
and with several different grid - cylinder spacings. In the following sections,
each flow configuration case with different ($TI$, $L_{12}$) is referred to with
its corresponding denotation in figure \ref{fig1}b.

\section{Visualisation and determination of the interface} \label{sec:interface
determinination} We start with a comparison of the visualisation of the wake of
the cylinder in a background flow without (figure \ref{fig2}a) and with (figure
\ref{fig2}b) turbulence, thereby featuring the distinction between a TTI and
TNTI. First, in the near wake (say $x/d \lesssim 10$), the large-scale vortices
are more distinct in the case of the non-turbulent background, whilst the locus
of the vortices' positions in the turbulent background extends to a further
lateral distance from the wake centre-line ($y = 0$). This confirms the
observation of \citet[]{kankanwadi2022near} that the large-scale vortices of the
near wake (identified via the velocity field, not the scalar field) in a
turbulent background generally drift to further positions in the lateral ($y$)
direction than those in a non-turbulent background at the same $x/d$ location.
Second, the TTI is also characterized by a ``rougher'' boundary with the ambient
fluid, at both large and small scales. Large-scale (intermittent) lumps of fluid
from the wake are observed protruding into the ambient flow in the turbulent
background case (say at $x/d \approx$ 26 and 33 in figure \ref{fig2}b), which is
barely seen in the non-turbulent background case (figure \ref{fig2}a). It is
also noted that there are more finer scale structures embedded into the TTI,
which is likely a reflection of the interaction between the smaller scale eddies
in the ambient turbulence and the interface. The resultant crinkled interface
(see also figure \ref{fig4_added} for different TTI cases examined) is later
demonstrated to have a very different fractal dimension to the TNTI, quantifying
our observation here that the TTI is ``rougher'' than the TNTI. 
\begin{figure}
    \centering
    \includegraphics[width = \textwidth]{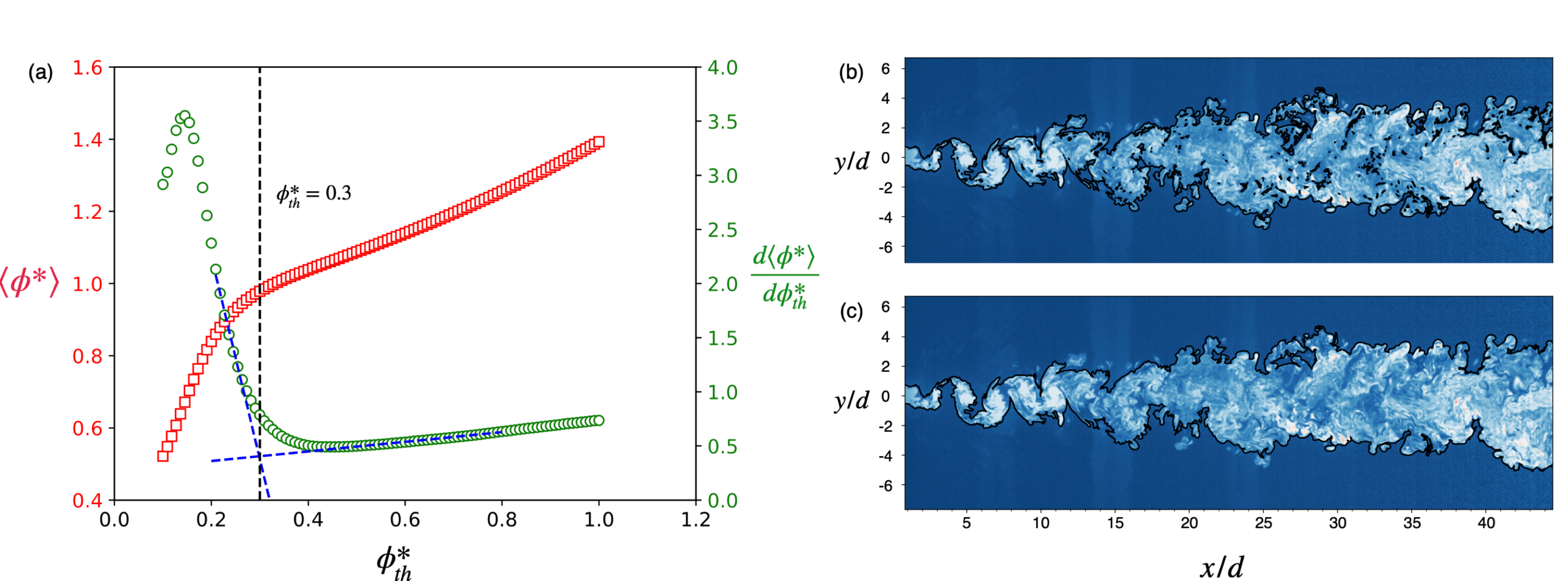}
    \caption{(a) Distribution of conditionally-averaged normalised light
    intensity $\langle \phi^* \rangle$ and ${\dt \langle \phi^* \rangle}/{\dt
    \phi_{th}^*}$ with respect to the threshold $\phi^*_{th}$. (b) Detected
    contours using $\phi_{th}^* = 0.3$. (c) Interface lines determined by
    selecting the longest continuous contours on both sides of the wake. All the 
    result are from case 1a.}
    \label{fig3}
\end{figure}

\begin{figure}
    \centering
    \includegraphics[width = \textwidth]{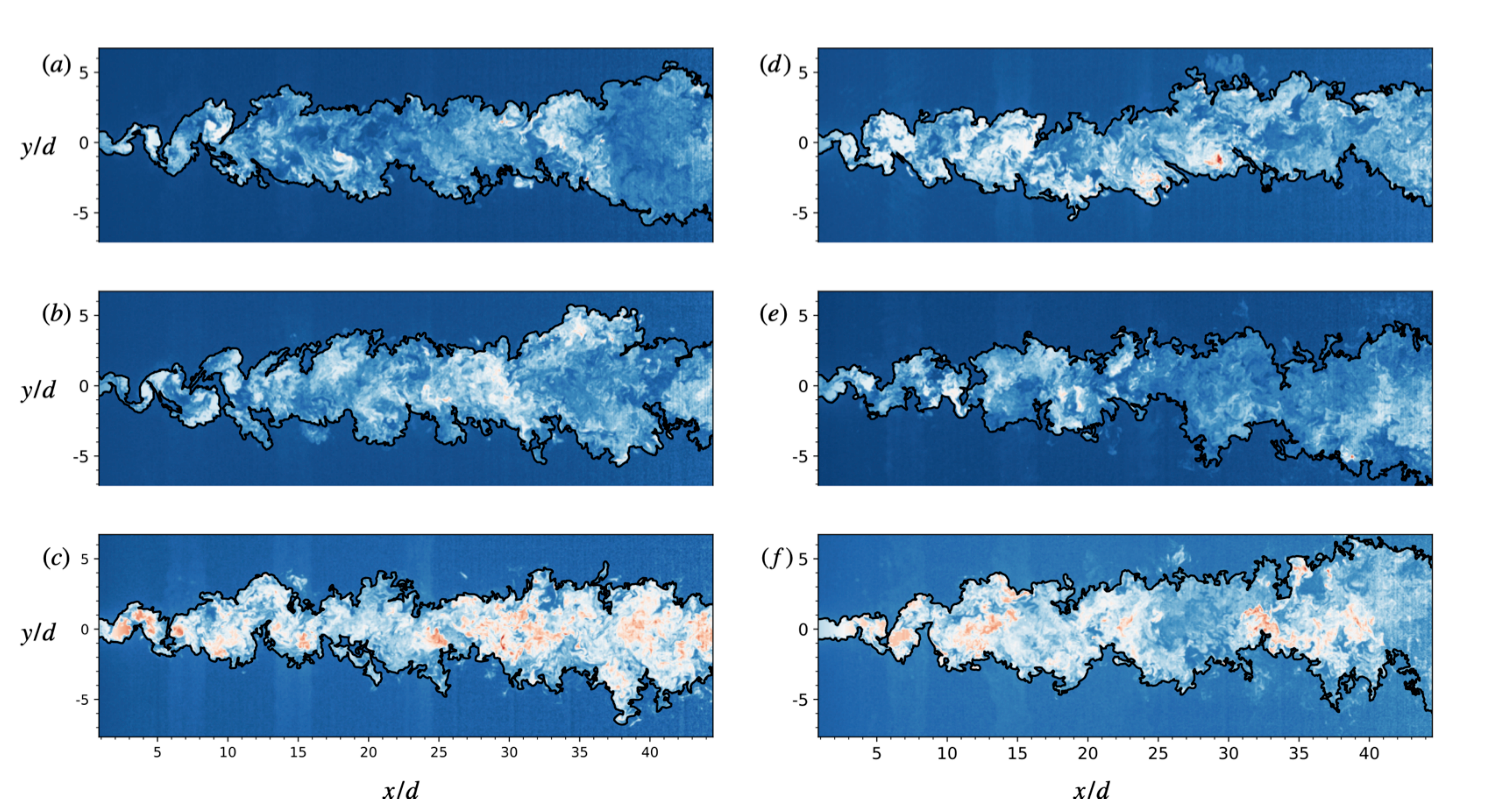}
    \caption{Typical interface of all TTI cases: (a) case 1b, (b) case 1c, (c)
    case 2a, (d) case 2b, (e) case 3a, (f) case 3b.  }
    \label{fig4_added}
\end{figure}

Before proceeding to examine the properties of the interfaces, we need to detect
their positions reliably. Both light intensity of the PLIF
images \citep[e.g.][]{ westerweel2009momentum, gampert2014vorticity,
mistry2016entrainment, kohan2022scalar} and the gradient of the light intensity
\citep[e.g.][]{silva2017behaviour, kankanwadi2020turbulent} can be used as the
threshold quantity in detecting the interface. As the measurement field of view
in the present study is relatively large (figure \ref{fig2}), we adopted the
light intensity method to avoid the streamwise variation of the threshold value
in using the light intensity gradient method. To account for the variation of the
light intensity along the streamwise direction in the PLIF images due to
mixing/out-of-plane transport (figure \ref{fig2}), the light intensity of each
image at each $x$ position is first normalised by its time-averaged mean value
at the same $x$ position along the wake centre-line, i.e. $ \phi^*(x,y,t)=
\phi(x, y, t)/\overline{\phi}(x, y = 0) $ where the overbar denotes the average
over time (images). The resultant normalised images enable a single threshold
value to be set for the entire field of view for interface identification
purposes (see figures \ref{fig3}b \& c). 

In order to determine this threshold we
follow the method used in \citet{prasad1989scalar} who also used PLIF to
distinguish the wake from the ambient flow in a similar experimental
configuration to our present study. Specifically, for each experimental case
(i.e. each data point in figure \ref{fig1}b), a conditional average was taken on
the normalised light intensity $\phi^*(x,y,t)$ exceeding the given threshold
value $\phi_{th}^*$ which reads
\begin{equation}
    \langle \phi^* \rangle \equiv \frac{\sum 
    (\phi^*|\phi^*>\phi_{th}^*)}{N(\phi^*>\phi_{th}^*)} .
\end{equation}
The distribution of $\langle \phi^* \rangle$ with respect to $\phi_{th}^*$ for
the wake with a non-turbulent background is shown in figure \ref{fig3}a. As
expected, $\langle \phi^* \rangle$ increases rapidly for small values of
$\phi_{th}^*$, but there is a knee point of $\langle \phi^* \rangle$ with
respect to $\phi_{th}^*$. This corresponds to the value of the light intensity
that well demarcates the limit between the background level of $\langle \phi^*
\rangle$ and that in the wake. The gradient $\dt \langle \phi^* \rangle/ \dt
\phi_{th}^*$ is also plotted in figure \ref{fig3}a and a threshold value of
$\phi_{th}^* =0.3$ was determined with the aid of a linear curve fitting on
either side of the knee point. We applied this value to a number of sample
images and it gives a good indication of the position of the interface in the
flow. A typical example of the detected interface is given in figure
\ref{fig3}b. 

One may note that small occasional patches inside and outside the
wake, which result from detrainment, three-dimensional ``teacup handle''
topology, or engulfment \citep{westerweel2009momentum} are also identified. Note
that these over-captured patches are disconnected from the continuous interface
we are seeking, so we chose the two longest continuous isocontours corresponding
to the threshold criteria and finally we obtain the interfaces on both sides of
the wake (see figure \ref{fig3}c). Typical interface isocontours of all the TTI
cases determined using the same method with case-dependent threshold value are
displayed in figure \ref{fig4_added} which all exhibit well-defined interfaces
between the wake and the ambient fluids. Comparison of the various figures also
beautifully highlights the dependency of the TTI geometry on both $TI$ and
$L_{12}$ of the background turbulence, with clear visual differences across the
various cases examined. 

\begin{figure}
    \centering
    \includegraphics[width = \textwidth]{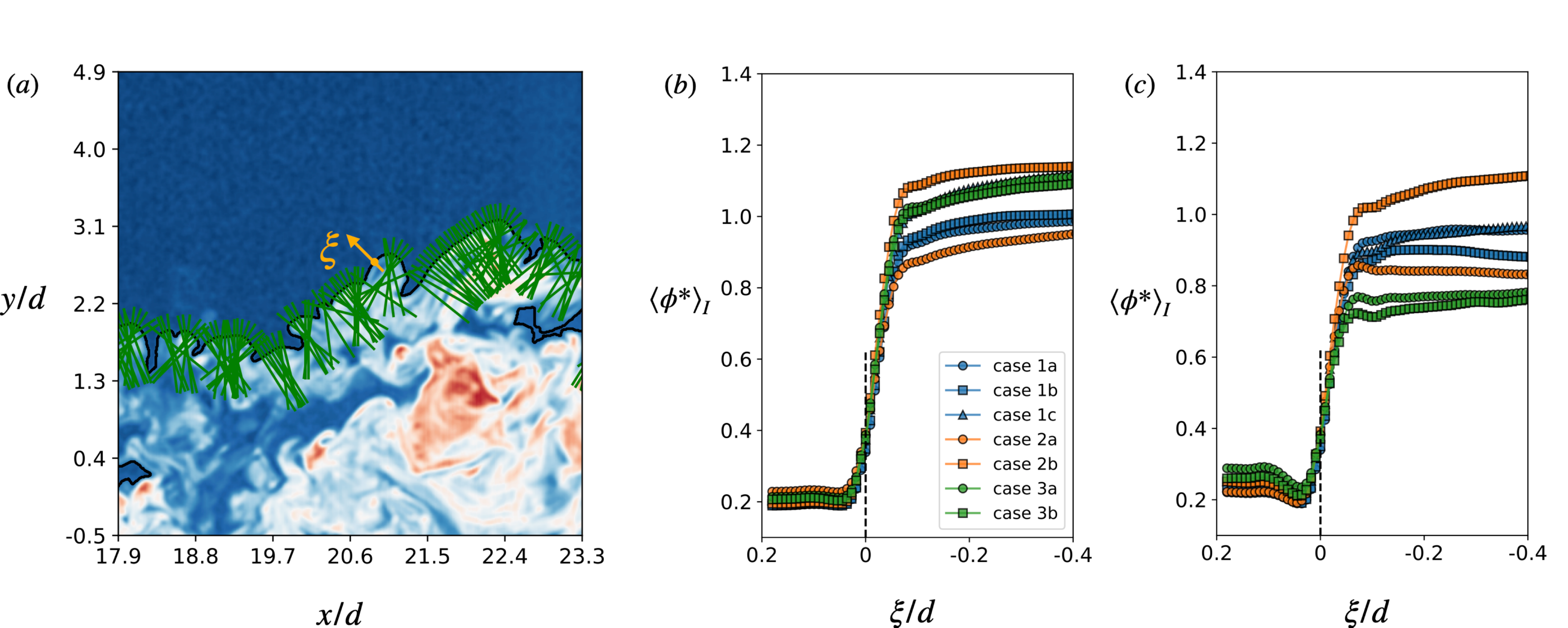}
    \caption{(a) Example of an interface segment of case 1a for the conditional
    average of the light intensity. The green bars normal to the interface show
    the path along which the conditional average is evaluated and the arrow
    points to the positive direction in (b) and (c). (b) Conditionally averaged
    light intensity across the interfaces ($\xi = 0$, indicated by the dashed
    line) in the normal direction of all the cases from $x/d = $ 5 to 40. (c)
    Conditionally averaged light intensity as in (b) but from $x/d = $ 38 to 40.}
    \label{fig5_added}
\end{figure}

Figure \ref{fig5_added} further confirms the validation of the interface
detection by showing the conditional average of the normalized light intensity
$\langle \phi^* \rangle_I$ across the detected interfaces in the
interface-normal direction (figure \ref{fig5_added}a). For all cases displayed
this calculation is performed from $x/d = $ 5 to 40 (figure \ref{fig5_added}b);
only the calculation paths that cross the interface once (the green lines in
figure \ref{fig5_added}a) are considered to avoid contaminating the result. The
clear jump of $\langle \phi^* \rangle_I$ across the interface ($\xi = 0$, where
$\xi$ is the ordinate on the interface normal, figure  \ref{fig5_added}a) for
all cases manifests as the rapid increase of the light intensity (associated
with the dye concentration) from the outer side ($\xi>0$) to the inner side
($\xi<0$) of the interface. We also confirm that the profiles of $\langle \phi^*
\rangle_I (\xi)$ retain a clear jump even when the conditional averaging is
performed locally in the far wake (e.g. $x/d = 38 - 40$ in figure
\ref{fig5_added}c). It is noted that the nominal interface position ($\xi=0$),
i.e. the isoline of the light intensity corresponding to the threshold value, is
slightly (about 0.2mm) dislocated from the start of the light intensity jump,
which is due to the marginally higher threshold value (between 0.3 and 0.4 for
all cases, figure \ref{fig3}a) than the background light intensity level (about
0.2) so as to differentiate the interface from the noise of the background. It
is interesting to see that on the inner side ($\xi<0$) of the interface, the
conditionally averaged light intensity is virtually a constant of order unity
which justifies the normalization we adopted for the PLIF images. By using a
separate PIV/PLIF combined measurement in the same configuration as the present
study (not shown for brevity), we have confirmed that the interface determined
from the fluorescent dye well matches the boundary of the wake determined from
the vorticity.

\begin{figure}
    \centering
    \includegraphics[width = \textwidth]{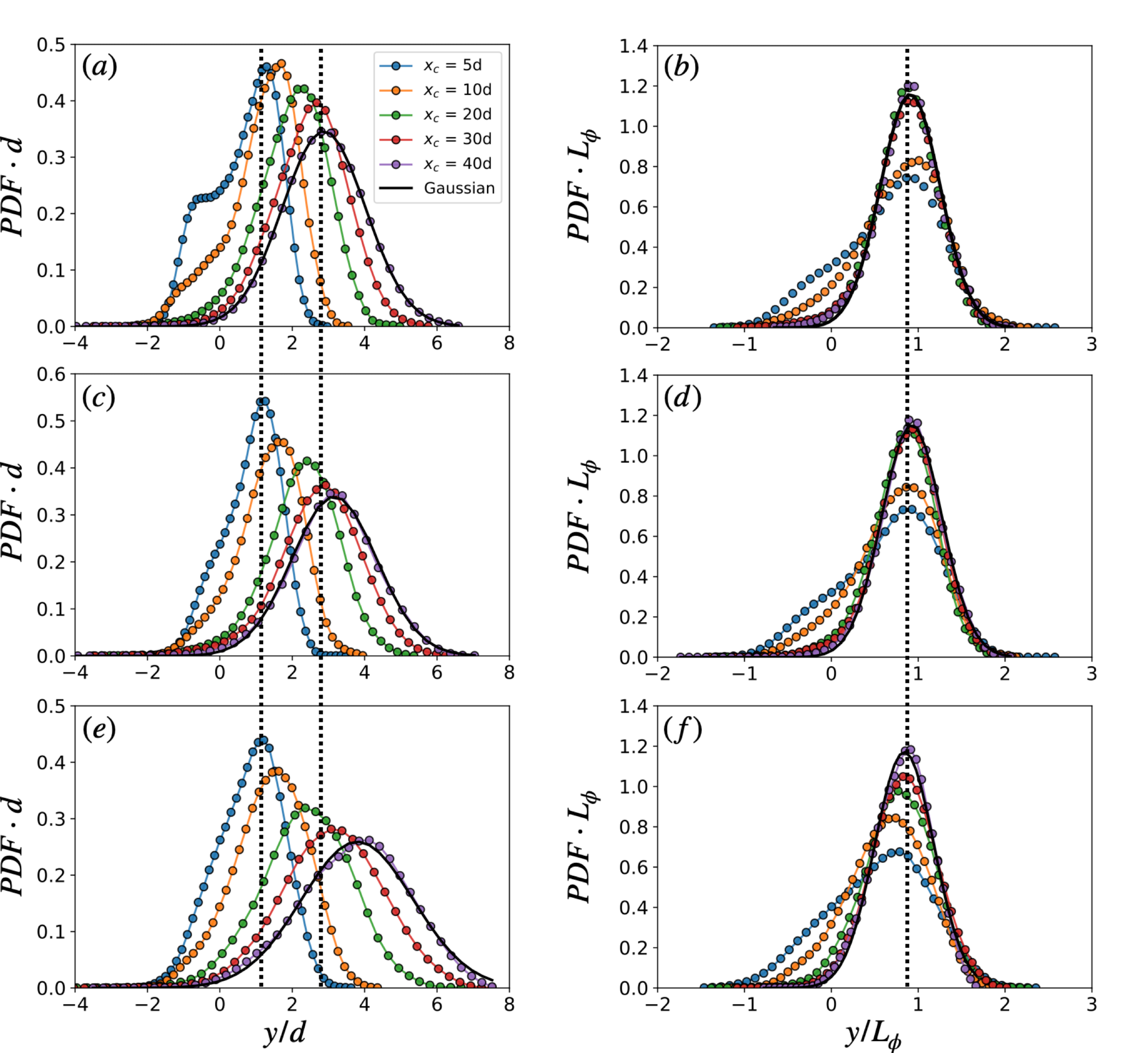}
    \caption{Streamwise development of PDFs of both TNTI and TTI position. (a,
    b) TNTI case 1a, (c, d) TTI case 2a, (e, f) TTI case 3a.}
    \label{fig4}
\end{figure} 

\section{Results and discussion}
\subsection{PDFs of TTI and TNTI position}

After the interface position was determined, the analysis proceeds first with
the downstream evolution of the PDFs of both TNTI and TTI position which are
examined at five different streamwise locations from very near to far away from
the cylinder, i.e. $x/d = $ 5, 10, 20, 30, and 40 (figure \ref{fig4}). For
presentational clarity, one typical case is displayed for each of the three
groups of figure \ref{fig1}b: figure \ref{fig4} (a, b) are plots of case 1a (the
TNTI case), and figure \ref{fig4} (c, d) and (e, f) are cases 2a and 3a
respectively (TTI cases). Both of the upper ($y>0$ in figure \ref{fig3}c) and
lower ($y<0$) interface lines are used in the calculation of the PDF, so a
negative value of $y/d$ in figure \ref{fig4} means the occurrence of a $y>0$ (or
$y<0$) interface on the $y<0$ (or $y>0$) side  at the examined $x/d$ position.
The PDF at a particular $x/d$ position was calculated within a streamwise strip
of extent $3d$ centred on $x_c$ as denoted in the figure. The $3d$ extent of
these strips is comparable to the largest integral length scale within the
background flow (see figure \ref{fig1}b), and enabled better statistical
convergence when computing the PDFs.

For the examined TTI cases (figure \ref{fig4}c, e), the modal peak of the PDF,
i.e. the most probable position of the interface which is very close to the mean
position of the interface, is roughly at the same position as that of the TNTI
case (figure \ref{fig4}a) at $x/d = 5$ (marked by the left dashed-line).
However, at $x/d = 40$ (marked by the right dashed-line), the position $y/d$ of
the modal location of the TTI is larger than that of the TNTI, especially when
the background turbulence intensity is high (figure \ref{fig4}e).
\citet{kankanwadi2022near} showed that in the near-wake region $x/d \leq 5$) the
wakes exposed to background turbulence were always wider on average than the
wake embedded in a non-turbulent background.
Our results show that in the near wake the modal position of the TTI is similar
to the TNTI and the reason that the mean wakes are wider for the TTI cases is
because of the diminshed contribution from the left tails of the PDFs (e.g. PDFs
for $x_c = 5$ in figure \ref{fig4}a, c, e), i.e. there are fewer instances of
the TTI crossing the centre-line than the TNTI. This observation is consistent
with the finding in \citet{kankanwadi2022near} that the mean position of the
centres of the von K\'{a}rm\'{a}n vortices for the TTI cases were further away
from the wake centre-line than those of the TNTI case at the same streamwise
position. Further, our results show that the increase in wake width in the
presence of background turbulence extends to the far wake, up to the $40d$
position examined in the present study, with the intensity of the background
turbulence seemingly the most important parameter in determining this enhanced
wake width. Later we will see that this average enhancement of the wake width
comes mainly from the contribution of the region closer to the cylinder which
then persists downstream.

\begin{figure}
    \centering
    \includegraphics[width = \textwidth]{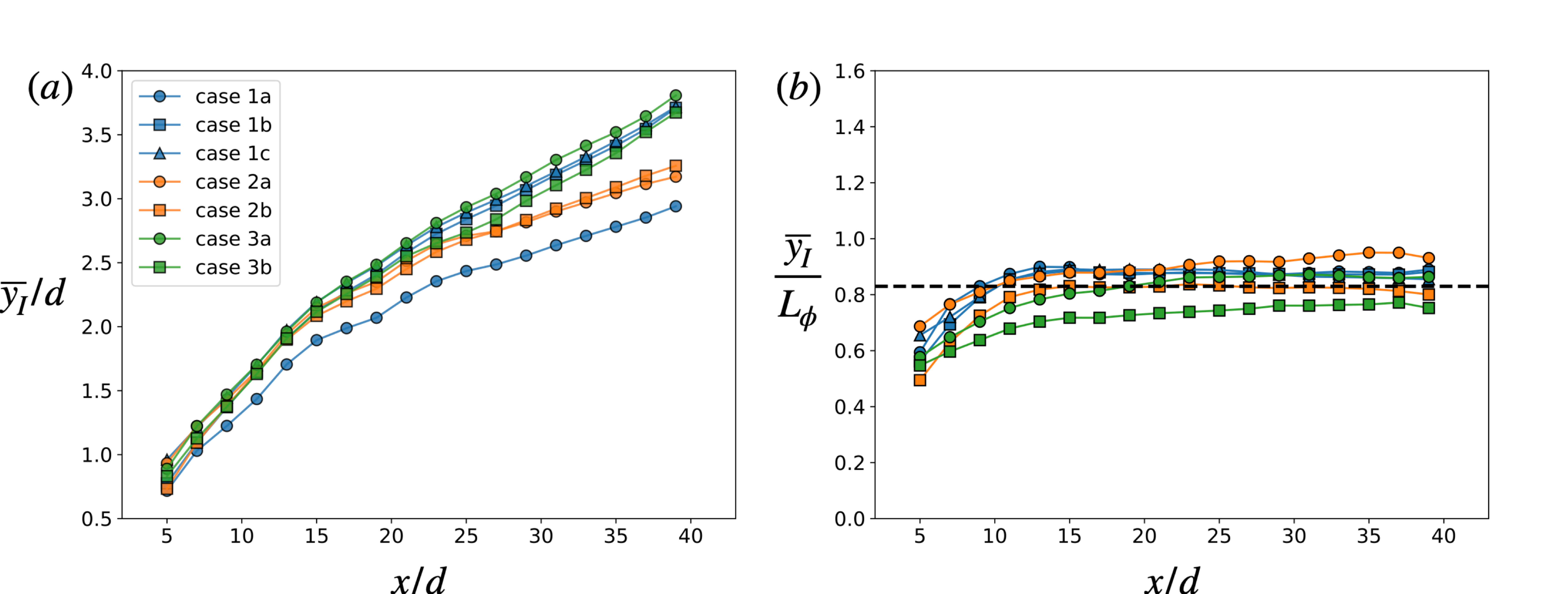}
    \caption{Streamwise distribution of the mean interface position (a)
    $\overline{y_I}/d$ and (b) $\overline{y_I}/L_\phi$ of all cases.}
    \label{fig5}
\end{figure}

It is noted that the TTI position PDFs for two cases with background turbulence
(figure \ref{fig4}c, e) are not Gaussian, with a negative skewness (not shown)
over all the examined $x/d$ range; a similar observation was also made by
\citet{kohan2022scalar} for the TTI position of an axisymmetric jet. The PDF of
the TNTI position is practically Gaussian at $x/d = 40$ (shown later in figure
\ref{fig7}) which has been widely reported in previous literature in
fully-developed regions of turbulent flows
\citep[e.g.][]{corrsin1955,da2014interfacial, mistry2016entrainment,
zhou2017related}. However, the TNTI PDF evidently deviates from a Gaussian
distribution at positions closer to the cylinder, especially at $x/d = 5$ and 10
(figure \ref{fig4}a) where heavier negative tails than for a Gaussian PDF are
displayed. These distinctly heavy negative tails reflect the high probability of
the interface appearing on the opposite side of the wake centreline ($y/d = 0$),
which is a manifestation of the strong large-scale ``meandering'' of the near
wake (see figure \ref{fig2}) because of the coherent vortices
\citep[e.g.][]{chen2016three,kankanwadi2022near}.

\citet{zhou2017related} found that the PDF of TNTI position in a turbulent,
axisymmetric wake scales with the wake width in the self-preserving region. Such
an observation is not made in the current study as shown in figure \ref{fig4}(b,
d, f) where the PDFs of both TNTI figure (\ref{fig4}b) and TTI (figures
\ref{fig4}d, f) position are normalised with the wake half-width $L_{\phi}(x)$
estimated from the mean profile of the light intensity $\overline{\phi}(x, y)$
of the PLIF images at the corresponding $x$ position. 
Here we
use the wake half-width determined from the mean scalar profiles instead of the
mean velocity profiles to represent the local characteristic length scale of the
wake, mainly because the velocity half-width is not well-defined for the cases
with a turbulent background. In particular, the inhomogeneity in the mean flow
of the background turbulence generated by the fractal grids (cases 1b, 1c and 2b
in figure \ref{fig1}) contributes to the difficulty of determining the boundary
of the mean velocity (and hence the maximum mean velocity deficit) of the wake.
In fact, $L_{\phi}(x)$ is shown to scale with the local velocity wake
half-width for the wake with a non-turbulent background flow (see Appendix A).
The normalised PDFs of both TNTI and TTI position for all cases assessed do not
collapse but exhibit an evident streamwise evolution. This is not unexpected as
the PDFs of either TTI or TNTI position in the current flow region are, as
discussed in the previous paragraph, heavily affected by the large-scale
coherent vortices and are not self-similar, as manifested by the heavy negative
tails at $x/d = 5 $ and 10. What is interesting to see is that the most probable
TTI and TNTI position do scale approximately with
the local $L_{\phi}$, which provides a straightforward way to estimate the most
probable position for both TNTI and TTI position, even though the PDFs are not
self-similar. 

\begin{figure}
    \centering
    \includegraphics[width = 0.6\textwidth]{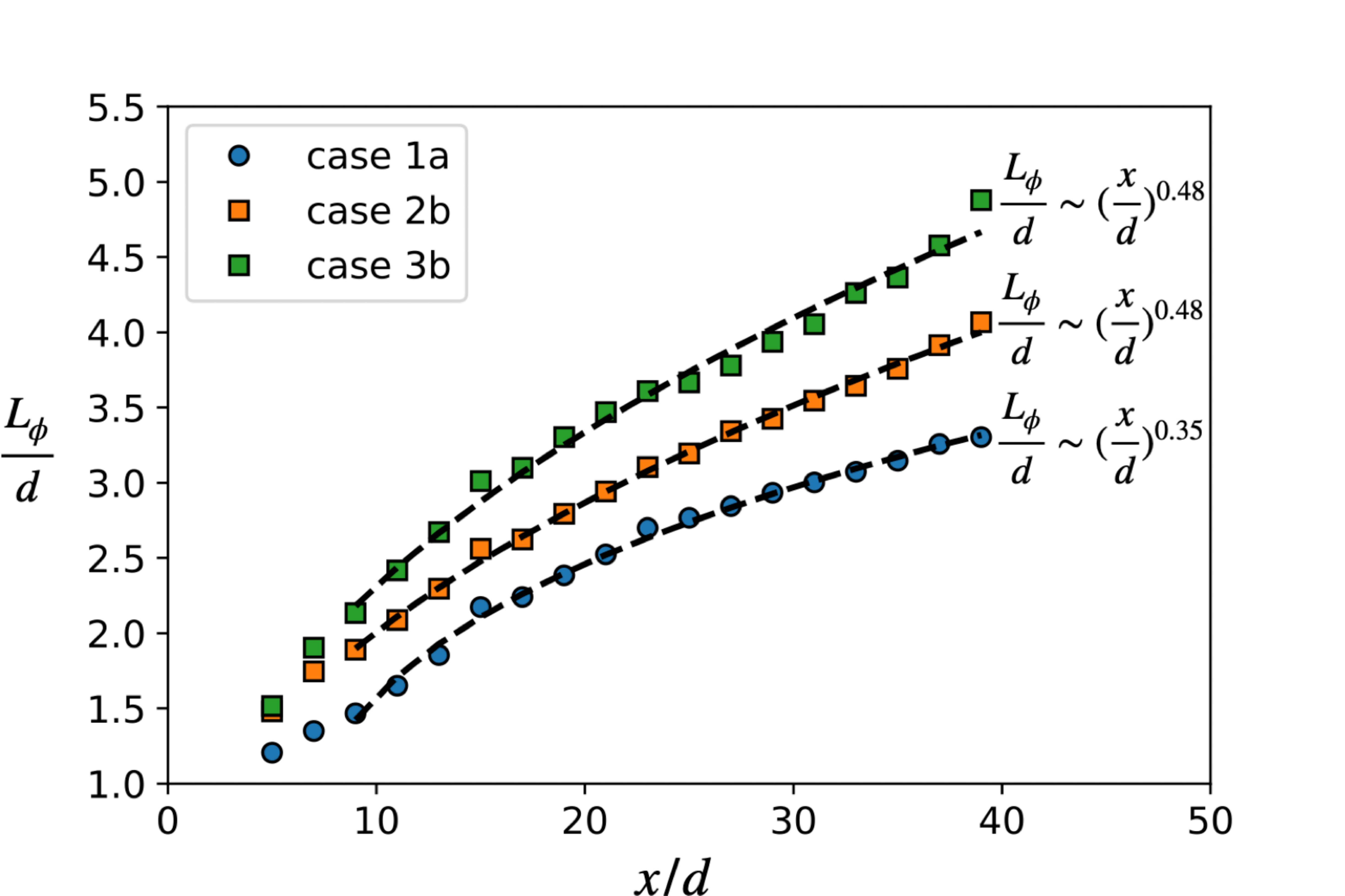}
    \caption{Streamwise distribution of the wake half-width $L_\phi$ with
    turbulent (cases 2b and 3b) and non-turbulent (case 1a) background flow.}
    \label{fig6}
\end{figure}

The coincidence of the modal peaks in figures \ref{fig4}(b,d,f) coupled to the
Gaussian-like nature of the PDFs for the further downstream locations suggests
that the mean position of the interface $\overline{y_I}(x)$ at different $x/d$
positions may scale with the local wake-half width. This is confirmed in figure
\ref{fig5} for both the TNTI case and all the TTI cases. Figure \ref{fig5}a
first compares the streamwise evolution of $\overline{y_I}(x)$ for both TNTI
(case 1a) and TTI cases scaled with the cylinder diameter $d$. It is clear that
all the TTI cases have a larger mean value of $\overline{y_I}$ than the TNTI
case at almost all $x/d$ positions, which is consistent with the observation in
figure \ref{fig4}(a, c, e). It seems the turbulence intensity is the dominant
parameter in determining the mean position of the TTI, as there is little
evident distinction between the TTI cases within groups 1 and group 2 in which
integral length scale is the major differentiating factor. What should be noted
is that the mean interface position at a particular $x/d$ location mainly
reflects the mass entrainment accumulated upstream of $x/d$, whilst the slope of
the curve $\dt \overline{y_I} / \dt x$ demonstrates the local entrainment rate
into the wake \citep{kankanwadi2022near}. It is found that in figure \ref{fig5}a
there is an apparent turning point of the slope of $\overline{y_I}(x)$ located
at $x/d \approx 15$ after which $\overline{y_I}(x)$ grows noticeably more slowly
than farther upstream, for both TNTI and TTI cases. It indicates that the
entrainment rate upstream of $x/d \approx 15$ is faster than after this
position. It is also noticed that in the flow region $x/d \lesssim 15$ the mean
interface position $\overline{y_I}(x)$ of the TTI cases grows almost linearly
and at a faster rate than the TNTI case; a similar observation was made by
\citet{kankanwadi2022near} in the flow region very close to the cylinder ($x/d
\leq 5$). It is thus concluded that the turbulence in the background promotes
spreading of the wake boundary mostly in the near wake region (say $x/d < 15$);
It is interesting to see that the turning point at $x/d \approx 15$ is almost
the same for all cases tested, regardless of whether there is a TNTI or TTI.
Although the physics underpinning the changes of the slope of
$\overline{y_I}(x)$ are still unclear, it is surmised that this transition
position may depend on the dynamics of the near-wake coherent vortices which
have been reported to be important for near-wake large-scale engulfment
\citep{kankanwadi2022near} and decays significantly from $x/d$ = 10 to 20 at a
similar Reynolds number \citep[e.g.][ and also the visualization in figure
\ref{fig2} ]{zhou2003three, chen2016three}. After this turning point, the growth
of the wake likely transitions from being large-scale engulfment-driven entrainment to
small-scale nibbling-driven entrainment

\begin{figure}
    \centering
    \includegraphics[width = \textwidth]{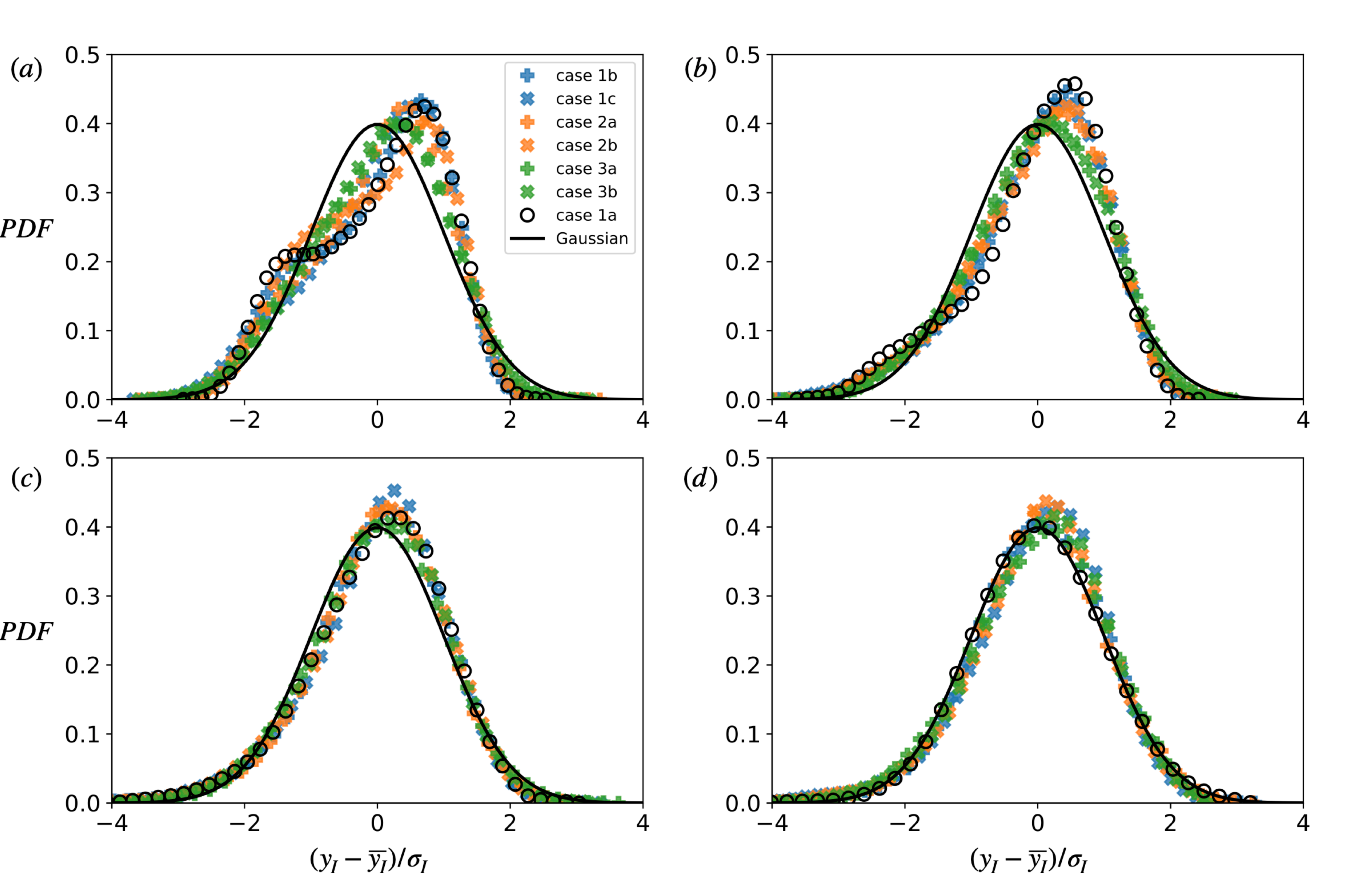}
    \caption{Comparison of centered PDFs of TNTI and all TTI cases at different
    $x/d$ positions. (a) $x/d = 5$, (b)10, (c)20, (d)40.}
    \label{fig7}
\end{figure}

When the mean interface position is scaled by $L_\phi$(figure \ref{fig5}b) all
$\overline{y_I}/L_\phi$ become approximately constant after an initial
development region ($x/d \lesssim  15$). This is consistent with the observation
in figure \ref{fig4} that the most probable position of $y_I$ scales with
$L_\phi$ which itself follows a power-law scaling while developing downstream
(figure \ref{fig6}). \citet{eames2011growth} developed a model that describes
how a wake spreads in a highly turbulent flow. They pointed out that for
two-dimensional bodies, the wake grows linearly with distance during the initial
development region (\citet{eames2011growth} called it `the ballistic regime')
until the wake width is comparable to the integral scale of the background
turbulence, beyond which the wake width grows diffusively with a scaling of
$\sim x^{1/2}$. Typical examples seeking a power-law scaling for $L_\phi \sim
x^\alpha$ are displayed in figure \ref{fig6}. After $x/d \approx 10$, the
scaling $L_\phi \sim x^{1/2}$ is indeed observed in almost all cases with a
turbulent background with the scaling exponent varying between $0.48 \leq \alpha
\leq 0.54$, except for two cases (case 1c and 2a in figure \ref{fig2} with a
scaling exponent of 0.64 and 0.23 respectively). It is noted that for the
non-turbulent background case (group 1a), the scaling exponent (0.35) is close
to 1/3, rather than the value of (1/2) expected based on the self-similarity
which is only achieved in the very far wake (say $x/d = 200$ in Ch. 4 of
\citet{tennekes1972first}).

We close the discussion of this section by a comparison between the centred PDF
of $y_I$ for all the examined TNTI and TTI cases (i.e. the PDF of
$(y_I-\overline{y_I})/\sigma_I$ where $\sigma_I$ is the standard deviation of
$y_I$) and a standard Gaussian distribution (figure \ref{fig7}), so as to
highlight the different extent to which the TNTI and TTI position PDFs deviate
from Gaussianity as $x/d$ varies. It is clear that very close to the cylinder
at $x/d = 5$ (figure \ref{fig7}a), the PDFs of both TNTI (case 1a) and all TTI
cases deviate from the Gaussian distribution significantly with evident negative
skewness, as is also seen in figure \ref{fig4}. 
Of note is that as $x/d$ increases, the negative skewness of the
TNTI position gradually reduces and the PDF becomes essentially Gaussian at $x/d
= 40$ (figure \ref{fig7}d), whilst the PDFs for all the TTI cases still deviate
from Gaussianity, although the skewness does reduce as $x/d$ increases. 
It is clear that for both
TNTI and TTI cases, the dynamics in the near wake (figure \ref{fig2}) are very
different from those farther downstream where the large-scale coherent vortices
have largely dissipated and the turbulence becomes fully developed.

Previous literature examining the TNTI position in turbulent jets
 \citep[e.g.][]{westerweel2009momentum,watanabe2014enstrophy,mistry2016entrainment}
 has also reported the slight deviation of the TNTI-position PDF from a Gaussian
 distribution with small, non-zero skewness. This implies that extreme
 interface-positions occur in one direction with a higher probability than in
 the other direction. Such large interface-position displacements from the jet
 centre are unlikely to be caused by the small-scale turbulent motions within
 the jet. We identify similar non-Gaussian TNTI-position PDFs for turbulent
 wakes (as well as for TTIs), e.g. $x/d = 5$ in figure \ref{fig7}a. Further
 downstream, where the large-scale coherent motions have largely dissipated, the
 TNTI-position PDF resorts to a Guassian distribution (e.g. $x/d = 40$ in figure
 \ref{fig7}d). Jets also experience a spatial evolution, with large-scale
 coherent motions embedded in the region near the jet exit \citep{ball2012flow},
 which subsequently decay with streamwise distance. Our results unequivocally
 support the conclusion that as the wake's coherent motions decay the
 TNTI-position PDF resorts from a non-Gaussian distrbution to Gaussianity. Given
 the similar spatial decay of the coherent motions in axisymmetric jets this
 lends support to the notion that non-Gaussian TNTI-position PDFs are associated
 to the presence of energetic coherent motions, regardless of flow type.

The different dynamics in the near and relatively far wake are believed to lead
to distinct geometrical features of the interfaces (TNTI and TTIs), which
encourages us to investigate the fractal dimension of the interfaces, and their
spatial evolution, in the next section.

\begin{figure}
    \centering
    \includegraphics[width = \textwidth]{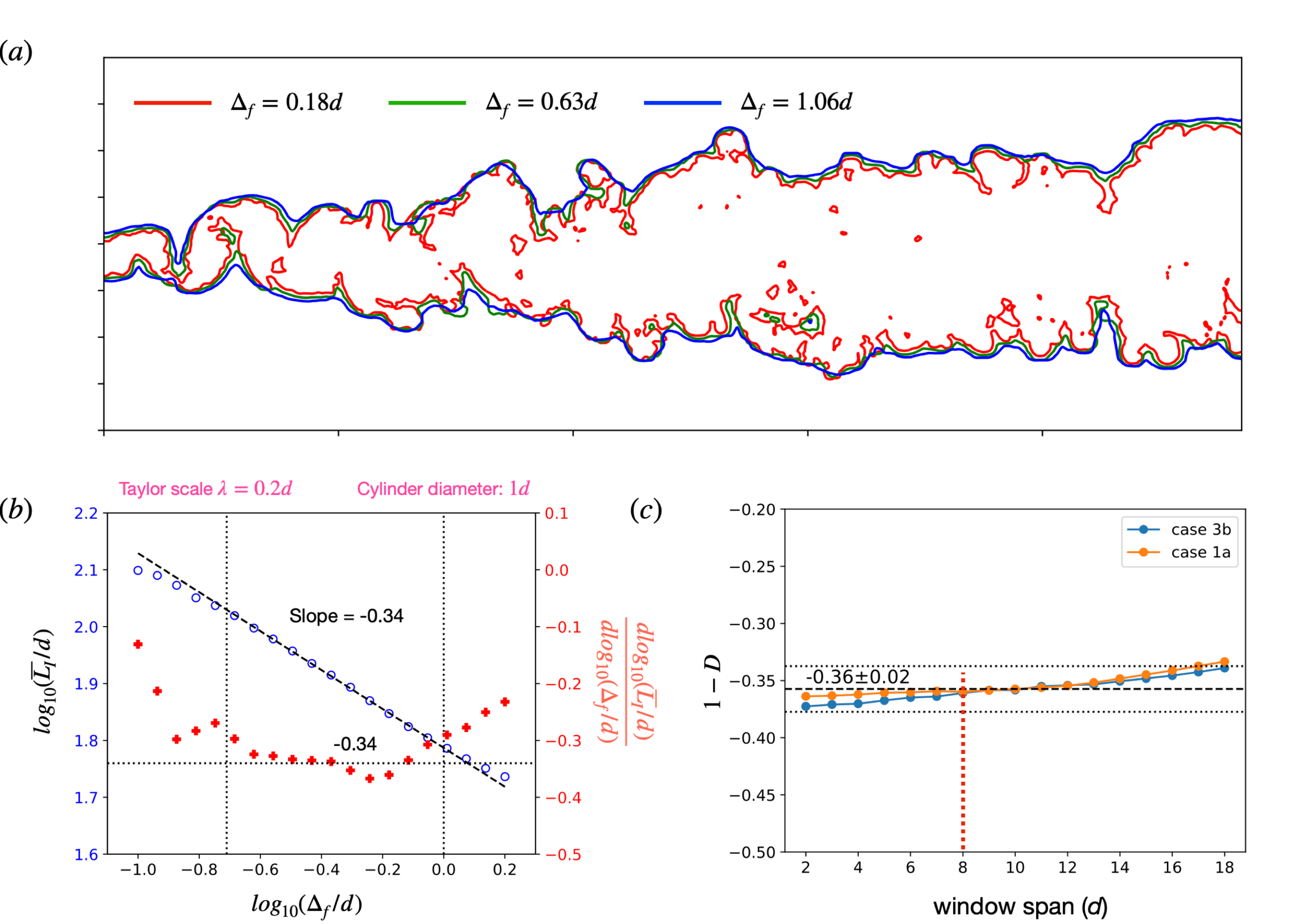}
    \caption{(a) Filtered interface with different filter
     scales (case 1a), (b) scaling of the length of the interface
     $\overline{L_I}$ in (a), and (c) fractal dimension of the interface
     obtained using different window widths for cases 1a and 3b at $x/d = 20$.
     The vertical dashed-line indicates the window width used for calculating
     the local fractal dimension of the interface.}
     \label{fig8}
\end{figure}

\begin{figure}
    \centering
    \includegraphics[width = 0.6\textwidth]{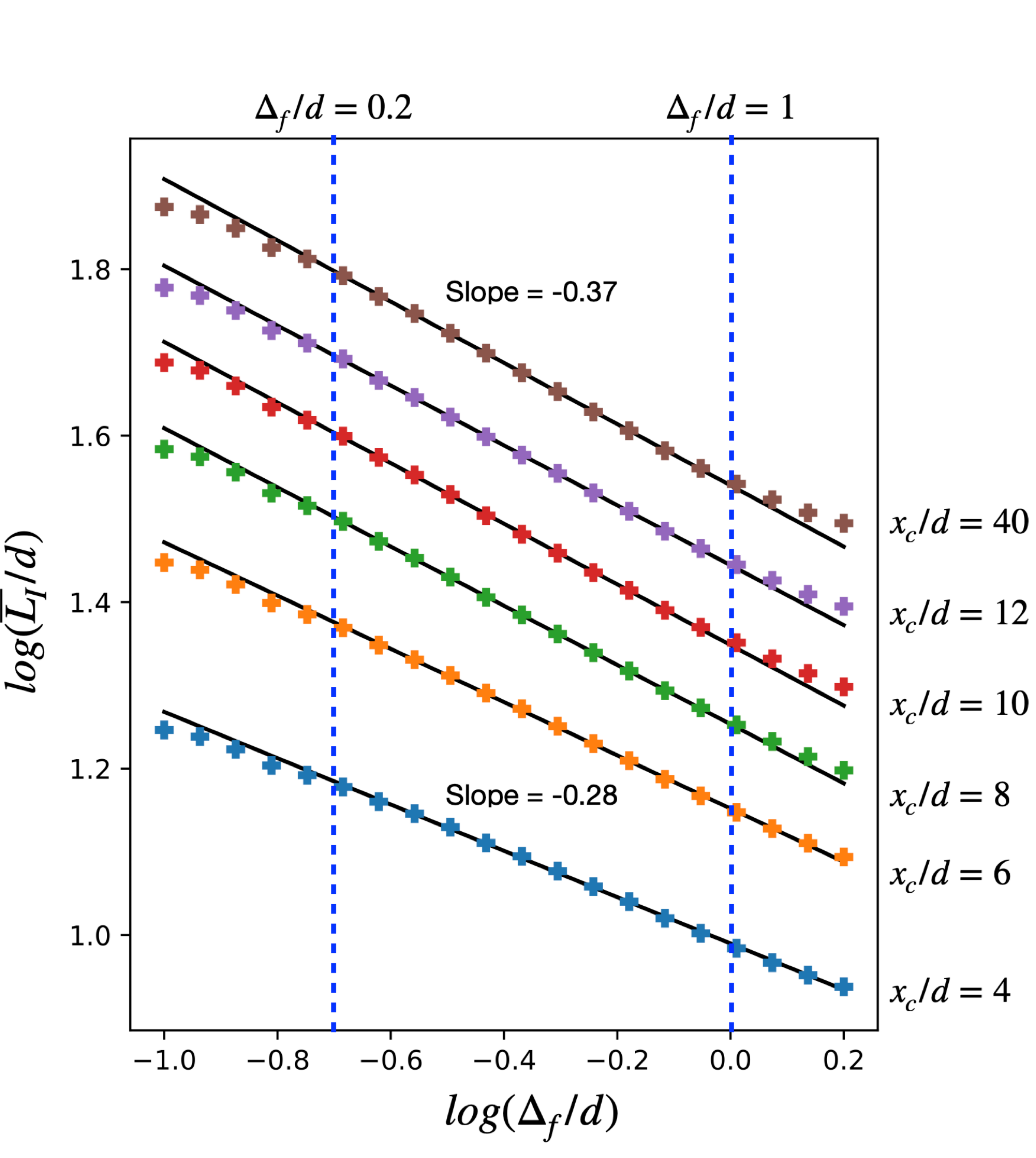}
    \caption{Scaling of the interface length of the TNTI case using window width
    of $8d$ at different streamwise $x/d$ positions.}
    \label{fig9}
\end{figure}

\subsection{Fractal dimension of the TNTI and TTI}
As explained in the introduction, the multi-scale self-similar geometric
features of the interface, either TNTI or TTI, can be described with fractal
analysis, which was first demonstrated by \citet{sreenivasan1986fractal}. The
length of a fractal ``line'' follows a power law with increased
resolution scale $r$, viz.,
\begin{equation}
    L_I(r) \sim r^{1-D}
    \label{eq:Li}
\end{equation}
where $D$ is the fractal dimension and has  been reported to be between 1.3 to
1.4 for a TNTI
\citep[e.g.][]{prasad1989scalar,de2013multiscale,abreu2022turbulent}, while for
the TTI the dimension is somewhat higher and an increasing function of the
turbulence intensity in the ambient flow
\citep{kankanwadi2020turbulent,kohan2022scalar}. However, these previous studies
focus on the TTI in the fully-developed region of a turbulent flow, where
\cite{kankanwadi2020turbulent} demonstrated that the turbulent length scale in
the ambient flow has little effect on the fractal dimension of the interface. In
the previous section, we have shown that the behavior of the interfaces are
substantially influenced by the strong organized motions in the near wake. As we
have measured multiple cases of TTIs with various levels of turbulent intensity
and integral length scales in the background flow, it is of interest to examine
the fractal dimension of these TTIs in the context of the streamwise decay of
the coherent vortices.

To obtain the fractal dimension of the interfaces, we adopt a `filtering method'
as used in previous studies
\citep[e.g.][]{de2013multiscale,kankanwadi2020turbulent,abreu2022turbulent}.
Specifically, each PLIF image is filtered with a box-filter of
different sizes $\Delta_f$ between $0.1d$ to $1.6d$; an interface is determined
after each filtering and the threshold of the light intensity used to identify
the interface is kept the same for all interface determination. Consequently,
the identified interface gets progressively smoother with resolution scale
smaller than the filtering size smeared (figure \ref{fig8}a). The lengths of the
interface corresponding to a particular resolution size (i.e. the filter size)
$L_I(r = \Delta_f)$ is obtained. Note that there are two lines on both sides of
the wake, whose length are calculated separately and both are included in the
ensemble to calculate the mean length of the interface line. Based on equation
(\ref{eq:Li}), $\log(L_I)$ has a linear relationship with $\log(r)$ when such a
scaling applies and the slope of the line (i.e. $1-D$, referred to as the
scaling exponent in the following text) is directly related to the fractal
dimension. Figure \ref{fig8}b displays a distribution of the mean
turbulent/non-turbulent interface length of all detected realisations with
respect to different filter sizes. In the scale range between $0.2d$ (close to
the Taylor microscale on the wake centreline at $x/d = 20$ of the TNTI case,
estimated from \citet{kankanwadi2022turbulent}) and 1$d$, a scale comparable to
the integral length there is a strong linear fit between
$\log(\overline{L_I}/d)$ and $\log(\Delta_f/d)$ with a slope of the fitted line
of -0.34. This yields a fractal dimension of $D = 1.34$ for the TNTI which
agrees well with the value in previous reports, e.g. 1.36 in
\citet[][]{prasad1989scalar}, 1.3 in \citet[][]{de2013multiscale} and $1.36 \pm
0.03$ in \citet{wu2020high}. 

To compute the local fractal dimension of the interfaces at different $x/d$ we
must choose a ``window'' covering a finite length of the whole interface; the
window span should be large enough to produce a good representation of the local
interface's fractality but small enough to ensure homogeneity over the
streamwise extent of the window and yielding good spatial resolution for the
fractal dimension's distribution (with respect to $x/d$). Figure \ref{fig8}c
shows the distribution of the scaling exponent ($1-D$), determined in the same
way as exhibited in figure \ref{fig8}b, with respect to different streamwise
window extents. Two typical cases are examined with the window centre set at
$x/d = 20$: the TNTI case 1a and the TTI case 3b which has the highest
turbulence intensity in the ambient flow (figure \ref{fig1}b). The value $1-D$
for both cases shows a weak increasing trend as the window span grows; there is
a narrow plateau between window spans of $7-11d$, displaying a reasonable value
of -0.36 \citep[e.g.][]{prasad1989scalar}. We therefore chose a window span of
$8d$ corresponding to the beginning of the plateau in the following study for
the best spatial resolution of the results.

\begin{figure}
    \centering
    \includegraphics[width = \textwidth]{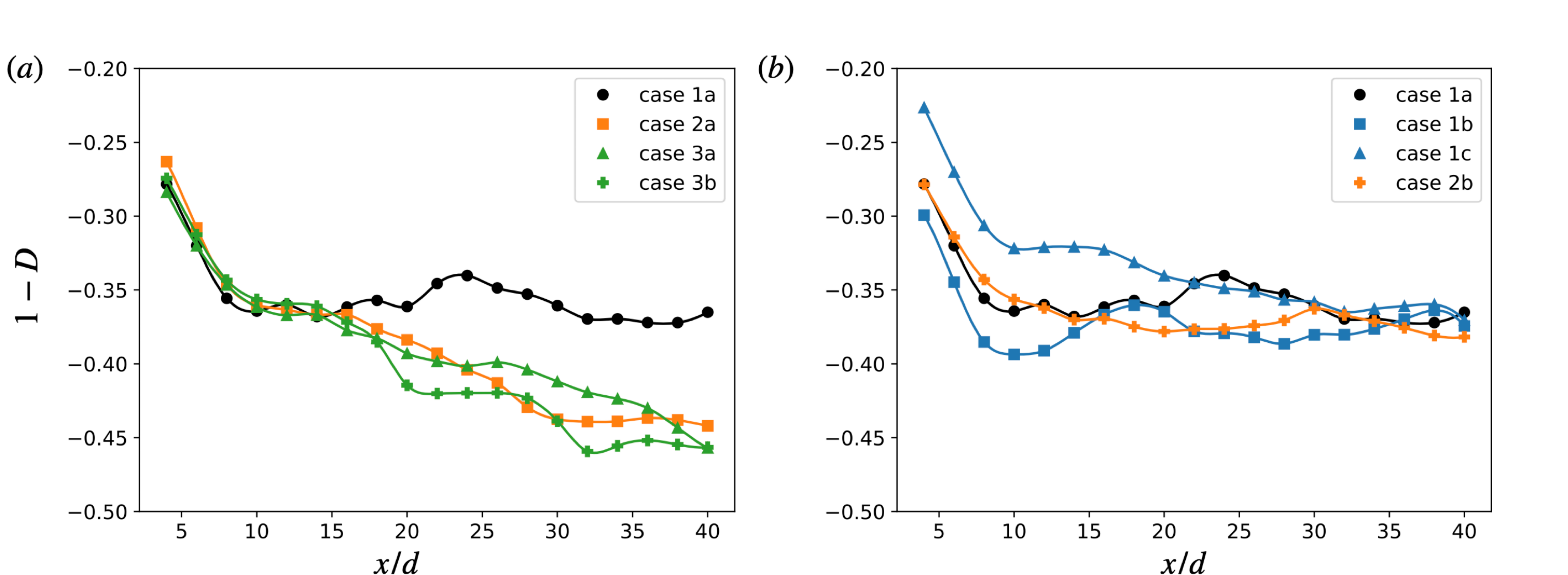}
    \caption{Streamwise distribution of fractal dimensions of TNTI and all TTI
    cases. (a) Effect of turbulence intensity, and (b) effect of integral length
    scale.}
    \label{fig10}
\end{figure}

Figure \ref{fig9} shows the scaling of the mean length of the filtered interface
of the TNTI case with respect to the scale of the filter at various streamwise
positions. In the figure, $x_c/d$ located between 4 to 40 is the centre position of the
examination window with span of $8d$. There is a well-defined scaling range
between $\Delta_f/d = $ 0.2 and 1 for all the examined positions, although the
scaling range is wider in the larger scale end for positions closer to the wake
generator. It is interesting to see that the slope of the fitted line ($ = 1-D$)
varies from -0.28 in the very near wake to an oft-reported -0.37 at $x_c = 40$,
indicating that there is indeed essential difference in the geometric features
of the interface in the near wake and the fully-developed downstream positions.
To explore the effect of the background turbulence on the fractal features of
the interfaces, we summarized the streamwise distributions of the scaling
exponent of all the measured cases in figure \ref{fig10}, in which the effect of
the background turbulence intensity and length scale on the fractal dimension is
respectively examined in figures \ref{fig10}a and \ref{fig10}b. 

In figure \ref{fig10}a, the streamwise distribution of the scaling exponent
$(1-D)$ of cases 2a, 3a and 3b, which are TTI cases with relatively small
integral scale and large turbulence intensity in the background flow (figure
\ref{fig1}b), are compared with that of the TNTI case (case 1a). The TNTI case
exhibits an approximately constant value around -0.36 in the region $x/d \gtrsim  
10$; the three TTI cases have similar distributions of $1-D$ to the TNTI case
before $x/d \simeq 15$ which interestingly corresponds to the position where the
wake spreading rate decreases evidently (figure \ref{fig5}a). After this $x/d$
position, the scaling exponent of the TTI cases continues to increase in
magnitude and reaches approximately $ -0.45$ at $x/d = 40$. The larger TTI fractal
dimension than that of the TNTI is also consistent with the observation of
\citet{kohan2022scalar} in an axisymmetric jet with a turbulent background. The
growing $(1-D)$ of the TTIs relative to the TNTI in the far field of the wake
indicates that the turbulence intensity in the background flow becomes gradually
essential in determining the fractal dimension of the interface in the positions
far from the wake generator. The increased fractal dimension in the far field of
the wake can also be observed in the visualisation in figure \ref{fig2}b, in
which the boundary of the wake becomes ``rougher'' (i.e. a larger fractal
dimension) as the flow proceeds downstream, with intermittent lumps and also
finer structures. These structures result from the interactions between the
eddies in the background turbulence and those of the wake. In the region closer
to the cylinder before $x/d \simeq 15$, the ambient turbulence does not
differentiate the scaling exponent of the TTIs from the TNTI. It implies that in
this flow region, which features the evolution of the strong von K\'{a}rm\'{a}n
vortices, the fractal nature of the interface is mainly determined by the
dynamics of the wake flow itself, at least when the scale of the energetic
eddies in the ambient flow is not overpowering (as is the situation of cases
cases 2a, 3a and 3b).

\begin{figure}
    \centering
    \includegraphics[width = 0.8\textwidth]{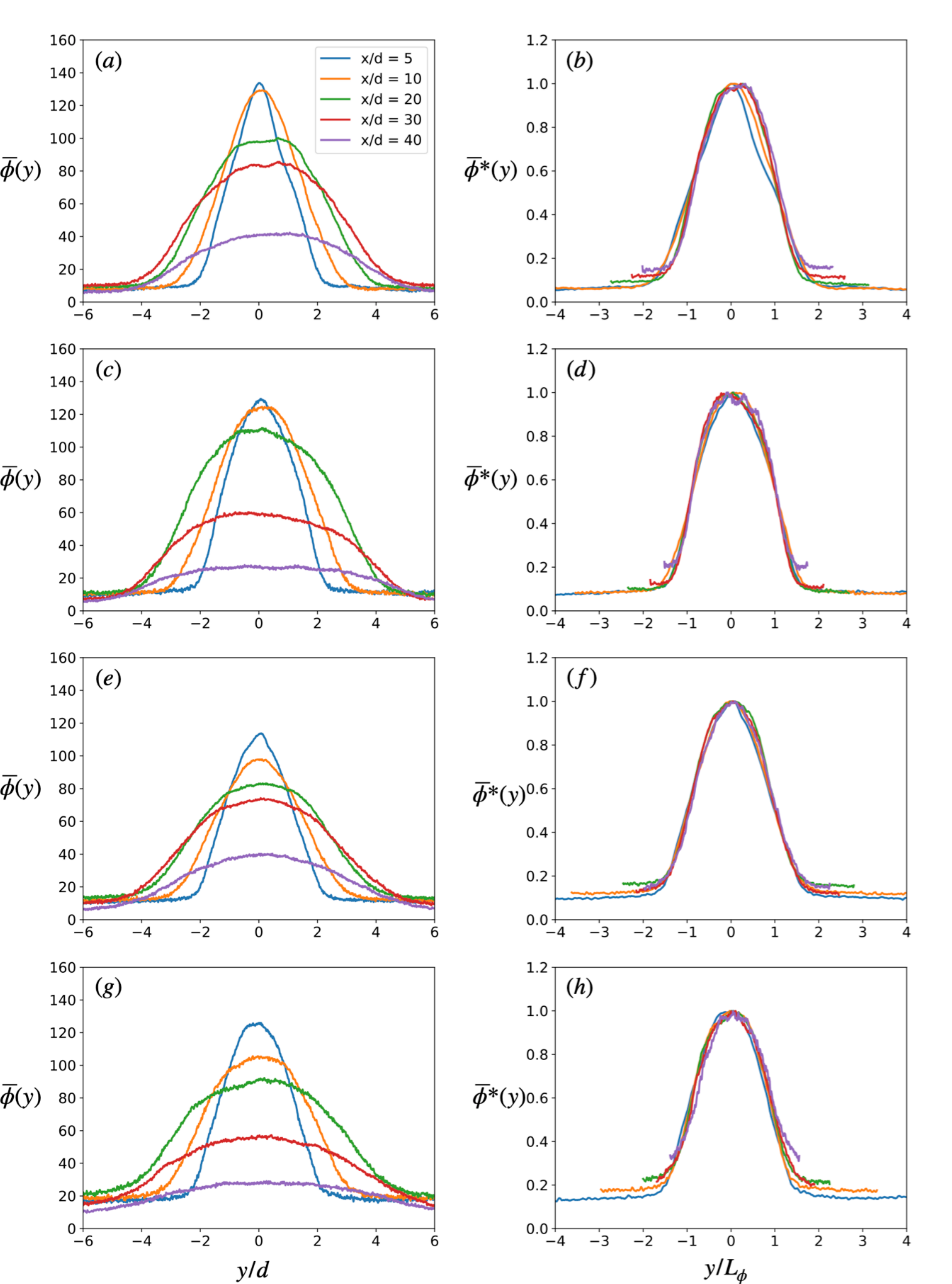}
    \caption{Profiles of mean light intensity of PLIF images of typical cases.
    (a, b) TNTI case 1a, (c, d) TTI case 1c, (e, f) TTI case 2a, (g, h) TTI case
    3b.}
    \label{fig11}
\end{figure}

In figure \ref{fig10}b, the TTI cases 1b, 1c which possess low turbulence
intensity and increasing integral length scale in the background flow are
compared with the TNTI case (case 1a); case 2b which has a large integral length
scale and also higher turbulence intensity is also added for comparison. In
contrast to the similar distribution of different cases upstream of $x/d \simeq
15$ in figure \ref{fig10}a, the scaling exponent distributions of the compared
cases show evident scatter in the upstream region but gradually converge to a
value $\approx -0.36$ in the downstream flow. TTI cases 1b and 1c differ from
the TNTI case 1a mainly in the integral length scale of the background flow,
their distinctive $1-D$ distribution in the upstream region indicates that the
integral scale of the background turbulence is of great importance to the
fractal dimension of the interfaces in this region. Compared to the TNTI case,
the TTI case 2b has both a higher turbulence intensity and a larger integral
length scale (figure \ref{fig1}b), and its distribution is not significantly
different from that of the TNTI case in both the upstream and downstream field.
It seems that there is a compound effect of the background turbulence intensity
and the integral length scale on the interface geometry. As a matter of fact,
such a combined effect was reported by
\citet{kankanwadi2020turbulent,kankanwadi2022near} in the same flow: in the
upstream field both the turbulence intensity and integral length scale in the
background flow act to enhance the entrainment rate into the wake whilst only
the turbulence intensity of the background turbulence is important in
suppressing entrainment in the downstream field.

To summarize the discussion of figure \ref{fig10}, generally, both turbulence
intensity and integral length scale in the background flow have an effective
influence on the fractal dimension of the interface, and in different regions of
the flow a different parameter is in dominance. In the near wake, the integral
length scale is the more important parameter; as the flow develops downstream
with the coherent vortices degrading substantially, the effect of the integral
length scale weakens and the influence of the turbulence intensity gradually
prevails. This observation is consistent with the conclusion obtained in the TTI
entrainment studies of \citet{kankanwadi2020turbulent,kankanwadi2022near} that
integral length scale is the more important parameter in the near wake which
promotes the large-scale engulfment of the wake, whilst the turbulence intensity
suppress the small-scale ``nibbling'' in the far field where the integral scale
is of less relevance.

\begin{figure}
    \centering
    \includegraphics[width = 0.6\textwidth]{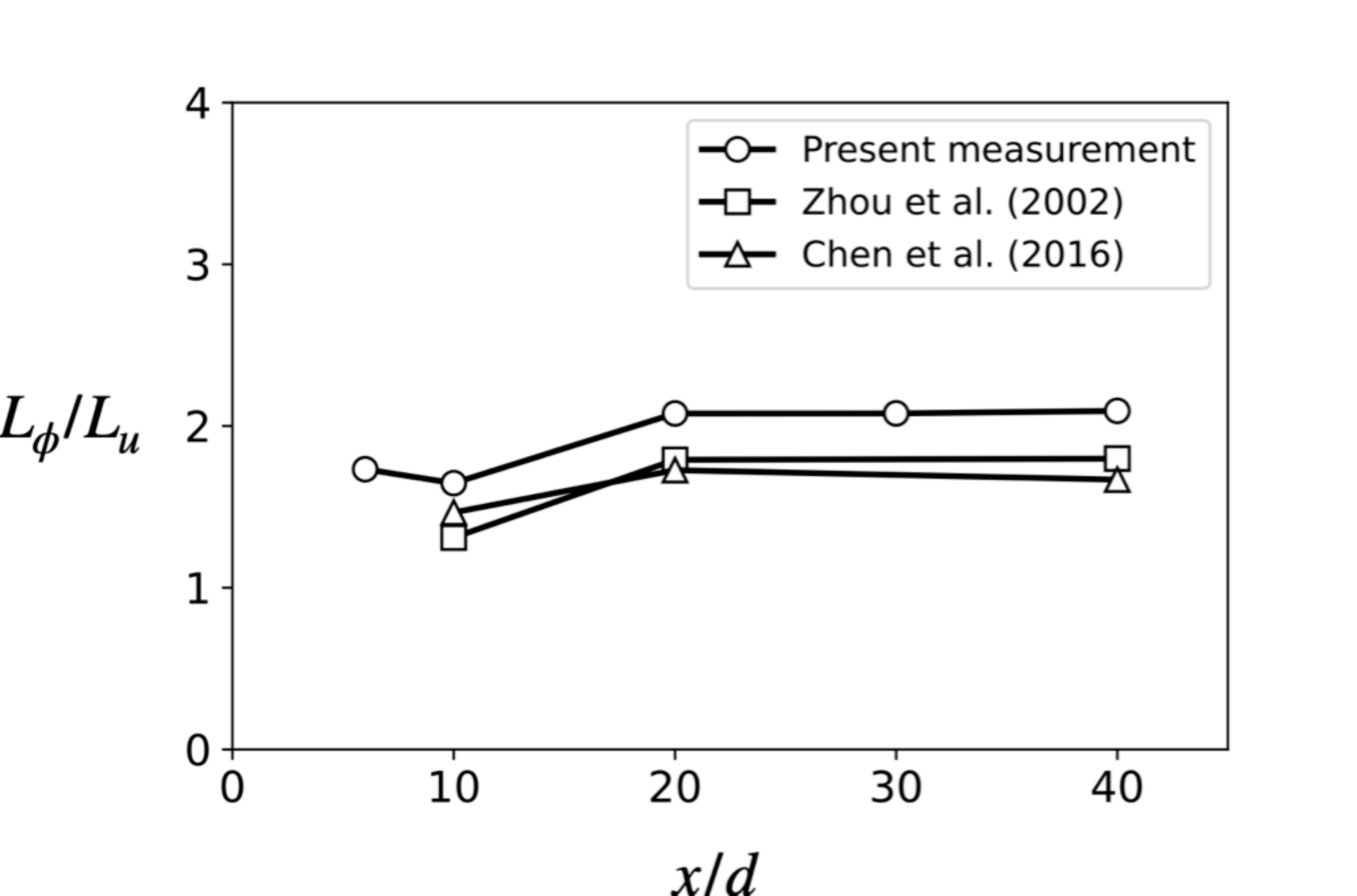}
    \caption{Ratio of scalar wake half-width $L_\phi(x)$ to velocity wake half-width $L_u(x)$ at
    different $x/d$ positions of a cylinder wake.}
    \label{fig12}
\end{figure}

\section{Summary and Conclusions}

We examined the spatial evolution of the geometry of the interface of a
turbulent cylinder wake from the near ($x/d = 5$) to the relatively far field
($x/d = 40$), in a turbulent background with various levels of turbulence
intensity and integral length scale (figure \ref{fig1}b). A PLIF experiment was
carried out to capture the interface between the wake and the turbulent
background flow. Attention was paid to the streamwise evolution of the geometric
properties of these TTIs and a TNTI reference case, including their PDFs,
scaling and fractality, in the context of the large-scale vortices gradually
diminishing in the wake.

Compared to the conventional TNTI, the TTI spreads faster towards the ambient
flow as the wake develops downstream, which is mainly due to the enhanced rate
of entrainment in the near wake \citep{kankanwadi2022near}. We find a transition
region of the interface spreading outwards at $x/d \approx 15$, after which the
interfaces spread at an evidently reduced rate (figure \ref{fig5}a). It is
conjectured that the different spreading rates before and after this transition
region are associated with the dynamics of the large-scale coherent vortices
which induce strong engulfment \citep[][also visualization in figure
\ref{fig2}]{kankanwadi2022near} and decay rapidly from $x/d = 10$ to 20 at
similar Reynolds numbers \citep[e.g.][]{zhou2003three,chen2016three}. After this
region, the mean position of the interfaces, including both TNTI and all TTI
cases, display a reasonable scaling with the wake half-width $L_\phi$(figure
\ref{fig5}b). $L_\phi$ is found to agree well with \citet{eames2011growth}'s
theoretical downstream evolution scaling of $(x/d)^{1/2}$ in a turbulent
background. It is interesting to see that this transition region is roughly the
same for both TNTI and TTI cases examined, suggesting that this transition
region is robust and not dependent on the turbulence in the background flow, at
least for the turbulence intensity and length scale range examined in the
present study.  

It is noted that the PDFs of both TTI and TNTI position are not Gaussian in the
near wake (especially for $x/d \lesssim  10$ ) with evident negative skewness
which reflects the ``deep-diving'' interface towards the wake central region due
to the strong engulfment by the coherent vortices at these locations (figure
\ref{fig2}). This observation is distinctly different from the oft-reported
Gaussian distribution of TNTI position at locations of fully-developed
turbulence in the absence of dominant coherent motions \citep[e.g.][and also the
Gaussian PDF of TNTI position at $x/d = 40$ in figure
\ref{fig7}d]{da2014interfacial, mistry2016entrainment, zhou2017related}. Note
that the PDFs of TTI position still depart from Gaussianity with a slight
negative skewness even at $x/d$ = 40 (figure \ref{fig7}d), which agrees with the
observation of \citet{kohan2022scalar} of the TTI in a fully-developed
axisymmetric jet.

We found that the fractal dimension of the TTIs in the near and
relatively far wake are dictated by different parameters of the background
turbulence. Turbulence intensity induces a  higher fractal dimension of the
interface in the far wake. It is highly likely to be resultant from the
interaction between the ambient eddies and those of the wake near the interface,
which can be partly observed from the evident intermittent small-scale
structures on the interface of the wake in turbulence background in figure
\ref{fig2}b. The effect of the integral length scale is more appreciable in the
near wake region (figure \ref{fig10}b). 
\citet{kankanwadi2022near} shows
that the freestream turbulence with large integral length scale can
significantly increase the transverse location of the centres of the von
K\'{a}rm\'{a}n vortices in the near wake ($x/d < 5$) relative to the
centreline. Therefore we have reason to expect that the
energetic eddies of the background flow interact with the coherent vortices
in the near wake, in which only the energetic eddies in the background flow with
comparable length scale or turnover time would interact effectively with the
coherent vortices in the wake. This is likely the reason why background integral
scale is important in the near wake. Such large-scale interactions in the near
wake would not necessarily wrinkle the interface, as the small-scale interaction
does in the far wake, explaining why cases with larger integral scale do not
necessarily cause higher fractal dimension of the interface (figure
\ref{fig10}b). Such large-scale interaction would be expected to cause
large-scale oscillation or meandering of the wake, however which has been
demonstrated by \citet{kankanwadi2022near}.

Finally, some comments about future studies are worthwhile.
The
    interaction between the background turbulence and the wake is certainty
    three-dimensional. The anisotropy of the turbulent scales in the background
    could cause different levels of distortion of the interface in the streamwise
    and spanwise directions, which can lead to different behaviour of the
    fractality of the interface in the two directions. This is an interesting
    topic to be addressed in future studies, either via direct numerical
    simulations or by imaging the wake in planes of different $x/d$.



\vspace{0.5cm}
\noindent
{\bf Funding.} The authors would like to acknowledge the Engineering and
Physical Sciences Research Council for funding the work under grant no.
EP/V006436/1

\vspace{0.5cm}
\noindent
{\bf Rights assertion statement.} For the purpose of open access, the author
has applied a Creative Commons Attribution (CC BY) licence to any Author
Accepted Manuscript (AAM) version arising 

\vspace{0.5cm}
\noindent
{\bf Declaration of interests.}  The authors report no conflict of interest.

\vspace{0.5cm}
\noindent
{\bf Author ORCIDs.} 

Jiangang Chen https://orcid.org/0000-0002-0976-722X;

Oliver R. H. Buxton https://orcid.org/0000-0002-8997-2986.

\appendix
\section{Determination of $L_\phi$}
This appendix is added to show how the wake half-width $L_\phi$ of the scalar
field is determined based on the PLIF measurements and its connection with the
velocity wake half-width $L_u$.

The profiles of the typical cases of the time-averaged light intensity of the
PLIF images, $\overline{\phi}(y)$ at $x/d = 5 $ to 40 are show in figure
\ref{fig11}. It is noted that the \underline{mean concentration} of the
fluorescent dye in the flow field is quite low and thus the fluorescent response
is effectively a linear function of the dye concentration
\citep{crimaldi1997effect,vanderwel2014accuracy,baj2016plif};
$\overline{\phi}(y)$ thus can be treated virtually as the concentration of the
dye which is confirmed later in figure \ref{fig12}. For all the cases considered
(figure \ref{fig11}a, c, e, g), $\overline{\phi}(y)$ reasonably decays in
magnitude and spreads into a wider range as $x/d$ increases. Similar to the
definition of velocity wake half-width, the scalar wake half-width $L_\phi$ is
defined such that $\overline{\phi}(y = L_\phi) = 1/2\overline{\phi}(y = 0)$. It
is interesting to find that the streamwise evolution of $\overline{\phi^*}$
scales well with $L_\phi$ for all the TTI cases (figure \ref{fig11}d, f, h); for
the TNTI case (figure \ref{fig11}b) $L_\phi$ also works well for $x/d \geq 20$.
It seems the scalar field of the wake in a turbulent background becomes
self-preserving at a smaller $x/d$ position than that in a non-turbulent flow. A
similar observation for the velocity field was also made by
\citet{eames2011growth}.

To the best of our knowledge, there is no published theoretical
prediction for the similar behaviour of $L_\theta$ and $L_u$ in wakes. However,
it is reasonable to envisage that there must be an interlink between $L_\theta$
and $L_u$, since the distribution of the scalar is passively determined by the
velocity field. This is confirmed in the present measurement of the
non-turbulent background case shown in figure \ref{fig12}, in which the ratio
$L_\phi/L_u$ is observed to be approximately constant at $x/d \geqslant 20$.
Here $L_u$ is the wake half-width determined from the mean velocity profile of
our non-published PIV measurement of the cylinder wake without grids upstream.
Note that a similar result was obtained in the measurements of
\citet{chen2016three} and \citet{zhou2002turbulent} in the wake of a cylinder at
the same $x/d$ range, except that their passive scalar was represented by
temperature in the flow. The slightly larger value of the present measurement
could possibly be attributed to the different initial conditions to those of the
two references: in our experiment the dye is released from a hole in the rear
surface of the cylinder while the scalar (heat) in \citet{chen2016three} and
\citet{zhou2002turbulent} is injected from the shear layer of the wake by
electrically heating the cylinder; in addition, the Sc number for the
fluorescent dye (about 2500, see section 2) is much larger than the Pr number
(about 0.7) of heat in air, which can also cause the scalar to being diffused
distinctly \citep[e.g.][]{rehab2001streamwise}. The resemblance between our
measurements and those from \citet{chen2016three} and \citet{zhou2002turbulent}
confirms our expectation that the distribution of the mean value of the
fluorescent intensity is a reasonable representation of the distribution of the
mean scalar concentration.

\bibliographystyle{jfm}
\bibliography{jfm}
\end{document}